\documentclass{article}
\usepackage{spconf,amsmath,graphicx,algorithm,algorithmic,adjustbox,setspace}
\usepackage{booktabs}
\usepackage{makecell}
\usepackage{amssymb}
\usepackage{hyperref}
\usepackage{array,multirow,graphicx}
\usepackage{caption}

\title{Boosting End-to-End Multilingual Phoneme Recognition through Exploiting Universal Speech Attributes Constraints}
\name{%
\begin{tabular}{@{}c@{}}
Hao Yen$^{1}$, Sabato Marco Siniscalchi$^{1,2,3}$, Chin-Hui Lee$^{1}$
\end{tabular}}
\address{$^1$Georgia Institute of Technology, USA\\$^2$Kore University of Enna, Italy\\$^3$Norwegian University of Science and Technology, Norway}
\begin{document}
\ninept
\maketitle
\begin{abstract}
We propose a first step toward multilingual end-to-end automatic speech recognition (ASR) by integrating knowledge about speech articulators. The key idea is to leverage a rich set of fundamental units that can be defined “universally” across all spoken languages, referred to as speech attributes, namely manner and place of articulation. Specifically, several deterministic attribute-to-phoneme mapping matrices are constructed based on the predefined set of universal attribute inventory, which projects the knowledge-rich articulatory attribute logits, into output phoneme logits. The mapping puts knowledge-based constraints to limit inconsistency with acoustic-phonetic evidence in the integrated prediction. Combined with phoneme recognition, our phone recognizer is able to infer from both attribute and phoneme information. The proposed joint multilingual model is evaluated through phoneme recognition. In multilingual experiments over 6 languages on benchmark datasets LibriSpeech and CommonVoice, we find that our proposed solution outperforms conventional multilingual approaches with a relative improvement of 6.85\% on average, and it also demonstrates a much better performance compared to monolingual model. Further analysis conclusively demonstrates that the proposed solution eliminates phoneme predictions that are inconsistent with attributes.
\end{abstract}
\section{Introduction}
\label{sec:intro}
\vspace{-.1cm}
End-to-end (E2E) automatic speech recognition (ASR) has made significant advancements in recent years~\cite{Li2021}. However, one of the key research challenges yet to be addressed in ASR is building a language-universal engine for all spoken languages. One potential solution is to explore shared acoustic-phonetic structures among different languages to build a large set of acoustic models~\cite{Cohen1997,Schults1998,Bryne2000,Lee2013} that characterize all the phone units needed to cover all the spoken languages being considered. Such approaches are called multilingual E2E ASR models, which have gained significant interest due to their capability to recognize multiple languages without prior knowledge of the language involved, e.g., \cite{Watanabe2017,Zhou2021,Zhang2023}. However, there are several opposing factors to be considered when designing multilingual E2E models~\cite{Lin2009}. On the one hand, expanding a multilingual model to accommodate a wide array of languages offers the benefits of more comprehensive contextual understanding and a greater range of recording conditions. On the other hand, this scaling introduces its own set of risks, such as data impurities stemming from variations between source and target languages, which could adversely affect the acoustic model of the target language. In addition, the presence of imbalanced data from different languages can result in mingled acoustic characteristics and context mismatches, which can lead to significant degradation in models that are sensitive to context and are trained with diverse speech data from various language sources. Therefore, a primary challenge in multilingual acoustic modeling lies in striking a balance between leveraging the rich, diversified acoustic data from multiple languages and mitigating the negative effects of data impurity and language mismatches. This has led to a growing interest in exploring knowledge sharing among multiple languages so as to define a universal set of acoustic-phonetic units that work for multiple or even for all languages.

Articulatory attributes are a set of distinct features that describe how speech sounds are produced by the articulators in the mouth, and these features are universally shared across all languages. Characterizing languages based on attributes offers two significant advantages over high-level characterization. Firstly, these attributes are universal across all languages, meaning that fundamentally there is no requirement to expand the number of attributes or their models when extending additional target languages. Furthermore, regardless of the languages, they can be refined once new training data becomes available. In previous studies on attribute modeling, automatic speech attribute transcription (ASAT)~\cite{Lee2013} has been utilized with success as a key component for several attribute-based speech applications. In ASAT framework, various detectors are trained to generate a bank of speech attributes. These detectors can be useful in many speech related task such as mispronunciation detection~\cite{Li2016,Li2016Detecting}, spoken language recognition~\cite{Siniscalchi2012,Lin2018}, lattice rescoring~\cite{Chen2014}, and continuous speech recognition~\cite{Lin2007,Mitra2017,Mitra2018,Siniscalchi2014}. The inherent sharable properties of articulatory attributes across languages make them a promising choice for building a multilingual ASR system, and we thereby expect to extend the technique to an E2E multilingual scenario that can maximize knowledge-sharing and naturally allow universal recognition.


\begin{figure*}[ht!]
    \vspace{-.42cm}
    \centering
    \includegraphics[width=0.9\textwidth]{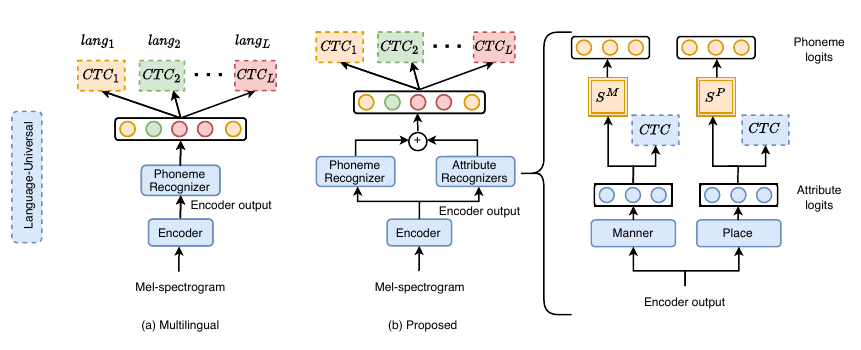}
    \caption{An overview of the traditional and proposed approaches. In (a), a multilingual approach predicting all the phoneme tokens for all languages. In, (b), our proposed system contains two articulatory attribute recognizers for manner and place of articulation respectively, that are shared across languages. The attribute recognizers will produce attribute logits that will later be transformed to phoneme logits with an attribute-to-phoneme mapping. $S^C$ stands for the articulatory attribute-to-phoneme mapping matrix for a specific attribute category $C$.}
    \vspace{-.4cm}
    \label{fig:proposed}
\end{figure*}

In this work, we present our initial step to build a multilingual E2E ASR by exploiting universal speech attributes. We construct a universal inventory for fundamental attribute units, and model articulatory characteristics across all languages, ensuring the universal attribute models to be rich in knowledge and capable of making accurate predictions. Furthermore, the auxiliary attribute recognitions serve as supplementary tasks that offer the benefit of learning a more robust representation and simultaneously make it possible to perform universal attribute recognitions, which can later be used for producing higher-level transcriptions, such as words. To connect attributes to phonemes, we present a precise deterministic mapping strategy that accurately projects output logits from various attributes to phoneme logits. Since the mapping matrices are pre-defined and non-trainable, which is different from ~\cite{Lee2023}, it can be viewed as a strict articulatory constraint imposed on the final predictions. We will discuss the differences in a later section and demonstrate that these added knowledge-based constraints have the advantage of reducing the possibility of inconsistent brute-force decoding results and improve recognition accuracy. Finally, it should be pointed out that Lee et al.~\cite{Lee2023} consider only two target languages and train both separately with specific target language data; whereas our proposed method is applied to six languages, namely English, Kyrgyz, Dutch, Russian, Swedish, and Tatar, without any prior knowledge about the target language involved.

\vspace{-.2cm}
\section{Related Work}
\vspace{-.1cm}
\label{sec:related}
\subsection{Multilingual Modeling}
Multilingual ASR can be divided into two main approaches. The first involves simply taking a union of the token sets (e.g., phoneme, character, subword, or even byte) of all languages to form the output token set and subsequently training an E2E model using all available data~\cite{Lin2009, Watanabe2017, Toshniwal2017}. As shown in Figure~\ref{fig:proposed}(a), such an E2E model enables it to recognize speech in any language that was included in its training data. However, sharing the output token space can inevitably present a new challenge. While it simplifies the process and optimizes cross-language sharing, it can also introduce ambiguity and confusion between languages during recognition~\cite{Li2021}.  The other approach aims to share some of the acoustic model parameters~\cite{Huang2013,Dalmia2018}. While the input and hidden layers are shared across all languages and can be considered as an universal front-end, the output layers, however, are not shared. Instead each specific language has its own output layer and perform phoneme recognition separately. An obvious limitation of such system is its total disregard for the phonetic correlations inherent to each language. In this study, we aim to provide a solution toward multilingual modeling without prior knowledge of the target language involved so we will focus on the first approach.

Recently, several studies have been conducted to tackle multilingual modeling using universal speech units. In~\cite{Li2020,Yan2021}, the authors incorporate a universal narrow phone set, called \textit{allophone} to build phone-based recognition models. Though \textit{allophone} can be considered as fundamental units, the total size of units is not compact and thus posing challenges in modeling when large amount of languages present. Moreover, the performance is dependent on the clarity of allophone-phoneme dynamics in the selected training languages.

\vspace{-.1cm}
\subsection{Articulatory Attribute Modeling for Speech Recognition}
An attribute modeling technique has been introduced to the task of speech recognition. M\"uller et al.~\cite{Muller2016} combined language feature vectors with articulatory attributes in an effort to improve low-resource ASR. Seven articulatory attribute classifiers were defined, and feature combinations for ASR was performed. Although a slight word error rate reduction is observed, the training is not end-to-end. In~\cite{Abraham2016,Qu2018}, multiple articulatory feature extractors are trained with CTC criterion for speech recognition. However, these methods require a second stage of feature integration and additional finetuning, which also deviates from our primary goal of building an end-to-end system. Li et al.~\cite{Li2019} investigated an efficient method of encoding multilingual transcriptions with articulatory representation and build a single attribute model based on E2E framework. Since some of the high-level units, such as phoneme and character, might be ambiguous, the modeling is rather restrictive and oversimplified. As a results, they observed that this method underperforms under most of the monolingual settings. However, it can significantly outperform usual E2E models under multilingual training, which implies that articulatory features can be shared across languages. In~\cite{Lee2023}, the authors incorporated articulatory information by embedding constraints in layer initialization. They built articulatory attributes predictors without explicit supervision and relied on the model itself to learn accurate predictions. By representing each token as a fixed-length encoding vector, they constructed an articulatory attribute projection matrix for each target language and expect the deep learning to finetune any incomplete, ambiguous or incorrect details. 

In contrast to \cite{Lee2023}, we train the attribute recognizers in a supervised manner, which can potentially produce more accurate attribute predictions, thus providing us with reliable indicators of the presence of any attribute. In addition, the number of recognizers can be naturally extended if more attributes are defined.  Regarding our attribute-to-phoneme mapping matrices, we deviate from the continuous scaling mapping technique employed in \cite{Lee2023}. Instead, we opt for a binary representation, where matrix entries are either 0 or 1. In this way, for a particular phoneme, a value of 0 signifies the absence of a specific attribute, while a 1 indicates its presence. Since most of the presences of attributes in each phoneme can be categorized in a binary representation, we contend that the deterministic binary mapping can reduce the ambiguity inherently associated with continuous scaling, which might harm the fine-grained attribute predictions produced by the attribute recognizers.

\vspace{-.1cm}
\section{Proposed Method}
\label{sec:proposed}
\subsection{Universal Attribute Inventory}
\label{sec:attr_invent}
\vspace{-1mm}
The set of speech attributes used in this work are acoustic phonetic features, namely, seven \textit{manner of articulation} classes (approximant, tap, fricative, affricate, nasal, stop, and vowel), and eleven \textit{place of articulation} (bilabial, labiodental, dental, alveolar, postalveolar, retroflex, palatal, velar, uvular, glottal, vowel). The inventory of universal attribute units adopted in this paper is shown in Table~\ref{tab:attributes}. Phonemes are categorized into each attribute based on the International Phonetic Association (IPA)~\cite{Smith2000} and PHOIBLE~\cite{phoible}, a large database of phone inventories for more than 2000 languages. As pointed out in~\cite{Tang2003}, a difficulty in using manner and place of articulation for ASR applications is that vowels and consonants cannot be mapped into a common linguistic space, because place of articulation has been differently defined for them. Therefore, in this study, vowels are categories in the same attribute.

\begin{table}[t!]
    \vspace{-.42cm}
    \centering
    \caption{Inventory of universal speech attributes used in this study.}
    \vspace{-.3cm}
    \label{tab:attributes}
    \begin{adjustbox}{width=\columnwidth}
    \begin{tabular}{c|l}
        \toprule
        \textbf{Category} & \textbf{Attributes} \\ \midrule \midrule
         Manner (M) &  approximant, tap, fricative, affricate, nasal, stop, vowel \\ \midrule
         Place (P) & \makecell[tl]{bilabial, labiodental, dental, alveolar, postalveolar, \\ retroflex, palatal, velar, uvular, glottal, vowel} \\ \bottomrule
    \end{tabular}
    \end{adjustbox}
    \vspace{-.3cm}
\end{table}

\vspace{-.1cm}
\subsection{Language-Universal Attribute and Phoneme Recognizers}
Figure~\ref{fig:proposed} (c) illustrates our proposed approach, which comprises a language-universal encoder along with a set of attribute recognizers, a phoneme recognizer, and a set of attribute-to-phoneme mapping matrices. The encoder first produces the input features and feeds them to both the phoneme recognizer and attribute recognizers. The phoneme recognizer aims to predict output phoneme tokens for each language.

A set of three attribute recognizers, for \textit{manner of articulation} and \textit{place of articulation} respectively, are placed to project input features to attribute logits. The attribute recognizers are trained as a predictor for articulatory attributes and optimized with full supervision via CTC criterion. Not only does the attribute recognizers abstract away from the phoneme recognizers, which contributes to the improvement in the multilingual acoustic modeling, the models also provides us the capability to predict universal articulatory speech attributes themselves. When trained with a large amount of available languages, these recognizers are expected to capture most of the distinct features that describe human sounds. Moreover, it has been shown that by joint modeling the articulatory and phonetic features, the performance of phoneme recognition can be improved accordingly~\cite{Mitra2017,Ji2022}.

\vspace{-.1cm}
\subsection{Attribute-to-Phoneme Mapping}
Following the language-universal attribute recognizers, two attribute-to-phoneme mapping matrices are used to project the attribute logits to the phoneme logits. Suppose there are $L$ training languages and each language $l$ has its own phoneme inventory $P_l$. A union phoneme inventory $P_\text{uni}$ can be created by enumerating all the phonemes appearing in the data for all languages:
\begin{equation}
    P_\text{uni}=\bigcup_{1 \leq l \leq L} P_l
\end{equation}
We can then construct an articulatory attribute-phoneme mapping matrix $S^C \in \mathbb{R}^{|C| \times |P_\text{uni}|}$, where $|P_\text{uni}|$ represents the number of phoneme tokens for the target language and $|C|$ represents the number of articulator attributes for a particular category of attribute, such as \textit{manner of articulation (M)}, and \textit{place of articulation (P)}, which means $C$ is one of $M$ or $P$. We can obtain $S^C = \{0,1\}^{|C| \times |P_\text{uni}|}$, describing the association of attributes and phonemes in each language $l$: Suppose the phoneme $p \in P_\text{uni}$ has the column index $j$ where $1 \leq j \leq |P_\text{uni}|$, attribute $a \in C$ has the row index $i$ where $1 \leq i \leq |C|$, if $p$ has the presence of attribute $a$, then $(i,j)$ entry of the matrix $S^C$ is set to 1, otherwise, it is assigned to 0. The attribute-to-phoneme mapping matrix is deterministic and fixed to accurately describe the relation between each attribute-phoneme pair. It also servers as a strong knowledge-based constraints to reduce any nonsense phoneme outputs.

As shown in Figure~\ref{fig:proposed}, the attribute-to-phoneme mapping aims to project attribute logits to phoneme logits. However, the attribute-to-phoneme projection does not yield a one-to-one correspondence, meaning multiple phonemes could potentially share identical logits after mapping. For example, as we mentioned in Section~\ref{sec:attr_invent}, vowels are categories into the same attribute \textit{vowel}, and therefore, the phoneme logits for all the vowels will simultaneously receive the same logits after projection. To address this limitation, it becomes necessary to introduce supplementary information for distinguishing among phonemes grouped within the same attributes. To address this, we combine the output logits of the attribute-to-phoneme projection with conventional phoneme logits to make the final predictions.

\begin{table}[t]
    \vspace{-.42cm}
    \centering
    \caption{Training corpora and size in hours for each language.}
    \vspace{-.3cm}
    \label{tab:dataset}
    \begin{adjustbox}{width=0.8\columnwidth}
    \begin{tabular}{llll}
        \midrule
        \textbf{Language} & \textbf{Corpora} & \textbf{Train (hrs)} & \textbf{Test (hrs)} \\
        \midrule
        English & LibriSpeech & 100.4 & 5.4 \\
        Kyrgyz  & CommonVoice & 2.6 & 2.0 \\
        Dutch   & CommonVoice & 11.5 & 7.0\\
        Russian & CommonVoice & 23.5 & 13.3 \\
        Swedish & CommonVoice & 2.1 & 2.0 \\
        Tatar   & CommonVoice & 11.9 & 4.6 \\
        \bottomrule
        \end{tabular}
        \end{adjustbox}
        \vspace{-.3cm}
\end{table}

\begin{table*}[th!]
    \vspace{-.42cm}
    \centering
    \caption{Phoneme Error Rate (PER(\%)) results for traditional models and proposed methods. Traditional methods include monolingual and multilingual models. Our proposed methods incorporated different categories of attributes, including \textit{manner of articulation}, and \textit{place of articulation}. The average PER (Avg)  for six languages is also reported.}
    \vspace{-1mm}
    \label{tab:results}
    \begin{tabular}{ c | c | c | c | c | c | c | c }
    \toprule
       \textbf{System}  & en & ky & nl & ru & sv & tt & Avg \\
       \midrule \midrule
       $\text{Mono}_\text{base}$ & 7.25 & 46.86 & 47.37 & 23.7 & 64.13 & 54.31 & 40.60 \\ \midrule
       $\text{Multi}_\text{base}$ & 16.45 & 41.13 & \textbf{44.82} & 28.94 & 45.79 & 56.84 & 39.00 \\
       $\text{Multi}_\text{attr}$ (ours) & \textbf{13.59} & \textbf{36.91} & 47.34 & \textbf{25.92} & \textbf{40.06} & \textbf{55.30} & \textbf{36.53}\\ \midrule
       Relative improvement (\%) & 17.39 & 10.26 & -5.62 & 10.44 & 12.51 & 2.71 & 6.85 \\ \bottomrule
    \end{tabular}
    \vspace{-3mm}
\end{table*}

\vspace{-.2cm}
\section{Experiments and Results}
\label{sec:exp}

\vspace{-.1cm}
\subsection{Datasets \& Experimental Settings}
Six languages are considered in the experiments, English (en) from LibriSpeech~\cite{Vassil2015} and Kyrgyz (ky), Dutch (nl), Russian (ru), Swedish (sv), and Tatar (tt) from the CommonVoice database~\cite{Ardila2020}\footnote{\url{https://voice.mozilla.org}}. Non-aligned phoneme transcription of each audio sample are obtained by running the open-source tool phonemizer\footnote{\url{https://github.com/bootphon/phonemizer}} on their corresponding text scripts. After processing all the transcriptions and removing some of the utterances that contain ambiguous phoneme tokens, the resulting training and testing dataset, and their corresponding sizes are shown in Table~\ref{tab:dataset}.

The Conformer architecture~\cite{Gulati2020} is investigated in this work. Our encoder first processes the input mel-spectrogram with a convolution subsampling layer and then with a number of conformer blocks. The Conformer block consists of three modules; a feed-forward module, a multi-head self-attention module, and a convolution module. The feed-forward module consists of a normalization layer and a linear layer with a Swish activation function~\cite{Ramachandran2018} followed by another linear layer. The multihead self-attention module is composed of a layer-normalization and multi-head self-attention with relative sinusoid positional embedding employed in Transformer-XL model~\cite{Dai2019}. The convolution module includes a layer normalization, a point-wise convolution layer with a gated linear unit (GLU) activation function~\cite{Dauphin2017}, and a 1-D depth-wise convolution layer. The depth-wise convolution layer is succeeded by a batch normalization layer, a Swish activation, and a point-wise convolution layer. We follow the \textit{small} Conformer setting, which incorporates 16 conformer blocks, each with 4 attention heads and the dimension is 144.

Phoneme and attribute recognizers are both one single-LSTM-layer decoder followed by a linear layer. The decoder dimension is 320. The models are trained with the Adam optimizer~\cite{Kingma2014} with $\beta_1=0.9$, $\beta_2=0.98$ and $\epsilon=10^{-9}$. A transformer learning rate scheduler~\cite{Vaswani2017}, with 10k warm-up steps and peak learning rate $0.05/\sqrt{d}$ where $d$ is the model dimension in conformer encoder, is used for adjusting learning rate. Each experiment is trained for 50 epochs with early stopping criterion.


\subsection{Results}
Table~\ref{tab:results} shows the phoneme error rate (PER) for several phoneme architectures in both monolingual and multilingual settings, including our system. The monolingual baseline system, denoted as $\text{Mono}_\text{base}$, is a model trained from scratch using only the target language data; whereas, $\text{Multi}_\text{base}$ refers to the baseline of multilingual modeling end-to-end trained on the union of all the phonemes from all target languages. On the one hand, the $\text{Multi}_\text{base}$ can improve the performance for some languages owing fewer training samples, for example, Kyrgyz and Swedish; on the other hand, it also creates confusion for resource-rich languages, such as English (7.25\% to 16.45\%), Russian (23.70\% to 28.94\%), and Tatar (54.31\% to 56.84\%), causing a degradation of the overall PER. Comparing the third with the second row in Table~\ref{tab:results}, we can see that our proposed $\text{Multi}_\text{attr}$ outperforms the $\text{Multi}_\text{base}$ for the majority of the target language with a relative improvement of 17.39\%, 10.26\%, 10.44\%, 12.51\%, and 2.71\% for English, Kyrgyz, Russian, Swedish, and Tatar, respectively.

Comparing with $\text{Mono}_\text{base}$, we observe significant improvements for Kyrgyz (46.86\% to 36.91\%), and Swedish (64.13\% to 40.06\%). For resource-rich languages, such as English, we see that, while we do not yet compete with the monolingual results in terms of PER, we effectively shrink the performance gap between the multilingual model and the monolingual model. It is also worth pointing out that our proposed $\text{Multi}_\text{attr}$ requires no prior knowledge regarding the target languages involved in the experiments. In contrast, $\text{Mono}_\text{base}$ needs to be trained and tested with specific language information.

\begin{figure}[t!]
    \centering
    \includegraphics[width=\columnwidth]{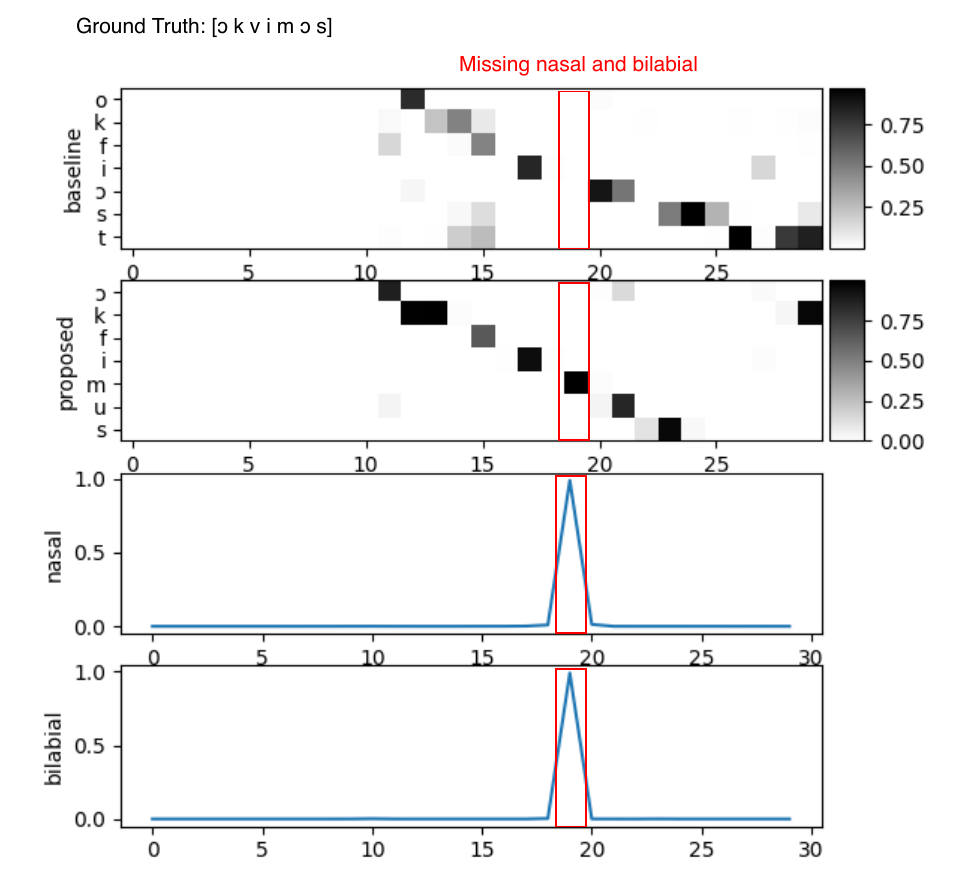}
    \caption{A qualitative example for utterance numbered 19557664 from the Swedish dataset. The above two plots are posteriograms of the first 30 frames for the utterance corresponding to the ground truth as shown above. The below two plots depict the activation curve for two attributes, namely \textit{nasal} and \textit{bilabial} from \textit{manner} and \textit{place of articulation} respectively.}
    \label{fig:example}
    \vspace{-.55cm}
\end{figure}

In order to gain more insights into the properties of our approach, we investigate into its ability to reduce inconsistent phonetic-acoustic evidence. Figure~\ref{fig:example} shows one qualitative example from the Swedish dataset. While the baseline system fails to predict the output phoneme [m] which has the presence of \textit{nasal} and \textit{bilabial} attributes from the ground truth, our manner and place recognizers clearly indicate the presence of those two events in the area where the mistakes occurs. The error is corrected by our solution since it has the ability to predict and integrate attributes information. 

\vspace{-.2cm}
\section{Conclusion}
\label{sec:conclusion}
We have presented the initial step to build a multilingual ASR system. Our results show that the proposed system can yield better performance than the conventional multilingual model in terms of phoneme error rate. Evaluated on six languages, we observed significant improvements for resource-limited languages, namely, Kyrgyz and Swedish. For languages with abundant resources, we successfully mitigated performance degradation. Further analysis has demonstrated that our proposed method can effectively eliminate inconsistencies with acoustic-phonetic evidence. While our system has yet to achieve the final goal of multilingual speech recognition, we contend that the remaining gap can be closed by integrating more attribute information, adding more target languages, and combining with a large pre-trained model, all of which are compatible with our framework.


\clearpage
\footnotesize
\bibliographystyle{IEEEbib}
\bibliography{refs}

\end{document}